\documentclass[amssymb,aps,prl,twocolumn,amsmath,showpacs]{revtex4}
\usepackage{epsfig}

\newcommand{\be}{\begin{equation}}
\newcommand{\ee}{\end{equation}}

\begin{document}

\title{Time and length scales in supercooled liquids}

\author{Ludovic Berthier}
\altaffiliation{Also at: Laboratoire des Verres, Universit\'e Montpellier II,
34095 Montpellier, France}
\affiliation{Theoretical Physics, University of Oxford, 1 Keble Road,
Oxford, OX1 3NP, UK}

\date{\today}

\begin{abstract}
We numerically obtain the first quantitative demonstration
that development of spatial correlations of mobility
as temperature is lowered
is responsible for the ``decoupling'' of
transport properties of supercooled liquids.
This result further demonstrates the
necessity of a spatial description of the glass formation 
and therefore seriously challenges a number
of popular alternative theoretical descriptions.
\end{abstract}

\pacs{64.70.Pf, 05.70.Jk}


\maketitle

Transport coefficients in liquids approaching the
calorimetric glass transition change by many orders of 
magnitude~\cite{review}.
It was discovered about a decade ago that conventional hydrodynamic relations
are not valid in supercooled liquids in the sense that various transport 
properties 
``decouple'' from one another~\cite{zphys,jl}. 
In a standard liquid the
viscosity, $\eta$, translational diffusion coefficient, $D$, 
and temperature, $T$, are linked by the hydrodynamic 
Stokes-Einstein relation, $D \eta \propto T$, which breaks down
for supercooled liquids. The product $D \eta$ can 
be orders of magnitude larger than its hydrodynamic
expectation at the glass transition, a puzzling observation 
which has received considerable interest in the last 
decade~\cite{zphys,jl,exp,theo,hetero}.

We have numerically obtained the first
quantitative
link between this decoupling phenomenon and the existence
of spatial correlations in the dynamics of supercooled liquids, 
also called ``dynamic heterogeneity''~\cite{hetero}.
We define here dynamic heterogeneity as 
expressing the fact that local dynamics of the liquid becomes 
{\it spatially} more correlated as $T$ decreases~\cite{juanpe}. This   
naturally implies the existence
of a wide distribution of relaxational time and length 
scales~\cite{juanpe,BG1}.
Decoupling results because different transport coefficients correspond 
to different averages over those broad distributions~\cite{theo}.

Increasing heterogeneity was previously reported
in a number of numerical works~\cite{glo},
but its direct relevance to transport was not established. 
On the other hand, experiments have quantified 
decoupling~\cite{zphys,exp,hetero},
but spatial correlations 
were only indirectly measured~\cite{hetero,explength} 
because they are presently inaccessible to scattering experiments 
since neutrons/light probe too small/large length scales, 
and no obvious temperature dependence was found~\cite{hetero,ed}.
The novelty of this work is therefore to reconcile in a direct 
manner two facets of the slow dynamics of supercooled liquids. 
Our results demonstrate the correctness of the widely shared belief 
that decoupling is most naturally interpreted in terms of heterogeneous
dynamics, a result which has been actively seeked in the last decade
and has deep theoretical implications.

Our argument shall be presented in three steps.
In large scale numerical simulations of a well-characterized
liquid model~\cite{walter}, we first quantify 
dynamic heterogeneity, then decoupling, to finally
establish in a third step the quantitative connection between
these two aspects. We finally discuss the important 
theoretical consequences of our findings.

We investigate the binary Lennard-Jones 
mixture proposed in Ref.~\cite{walter} with $N_A = 1097$ particles
of type $A$ and  
and $N_B= 275$ particles of type $B$ at density $\rho=1.2$.
The $N = N_A + N_B$ 
particles interact via a Lennard-Jones potential 
$V({\bf r}_{\alpha \beta}) = 4 \epsilon_{\alpha \beta} \left[ 
(\sigma_{\alpha \beta}/r_{\alpha \beta})^{12} - 
(\sigma_{\alpha \beta}/r_{\alpha \beta})^{6}
\right]$, with $\alpha, \beta = A, B$. Time, energy and length
are measured in units of $\sigma_{AA}$ and $\epsilon_{AA}$, 
and $\sqrt{m_A \sigma_{AA}^2 / \epsilon_{AA}}$, respectively.
Other parameters are $\epsilon_{AB} = 1.5$, $\epsilon_{BB} = 0.5$, 
$\sigma_{BB} = 0.88$, $\sigma_{AB} = 0.8$.
Newton equations are integrated via a leapfrog algorithm with 
time step 0.01. Velocity rescaling is used to thermalize the system.
We study the system at equilibrium 
for a wide range of temperatures, $T \in [0.42, 2.0]$. 
Since we measure fluctuations of local dynamical quantities, 
extremely long simulations are necessary to ensure not only
thermal equilibrium, but also that sufficient statistics 
is recorded. We use the parallelized
algorithm developed by Plimpton~\cite{plim}, 
and at all $T$ runs of length at least $100 \tau_\alpha$ are performed.
Our computer capabilities fix the lowest studied temperature.
Characteristic temperatures for this system are 
the onset of slow dynamics, $T_o \approx 1.0$, and $T_c \approx 0.435$, 
the location of the mode-coupling singularity in the analysis
of Ref.~\cite{walter}.

\begin{figure}
\begin{center}
\psfig{file=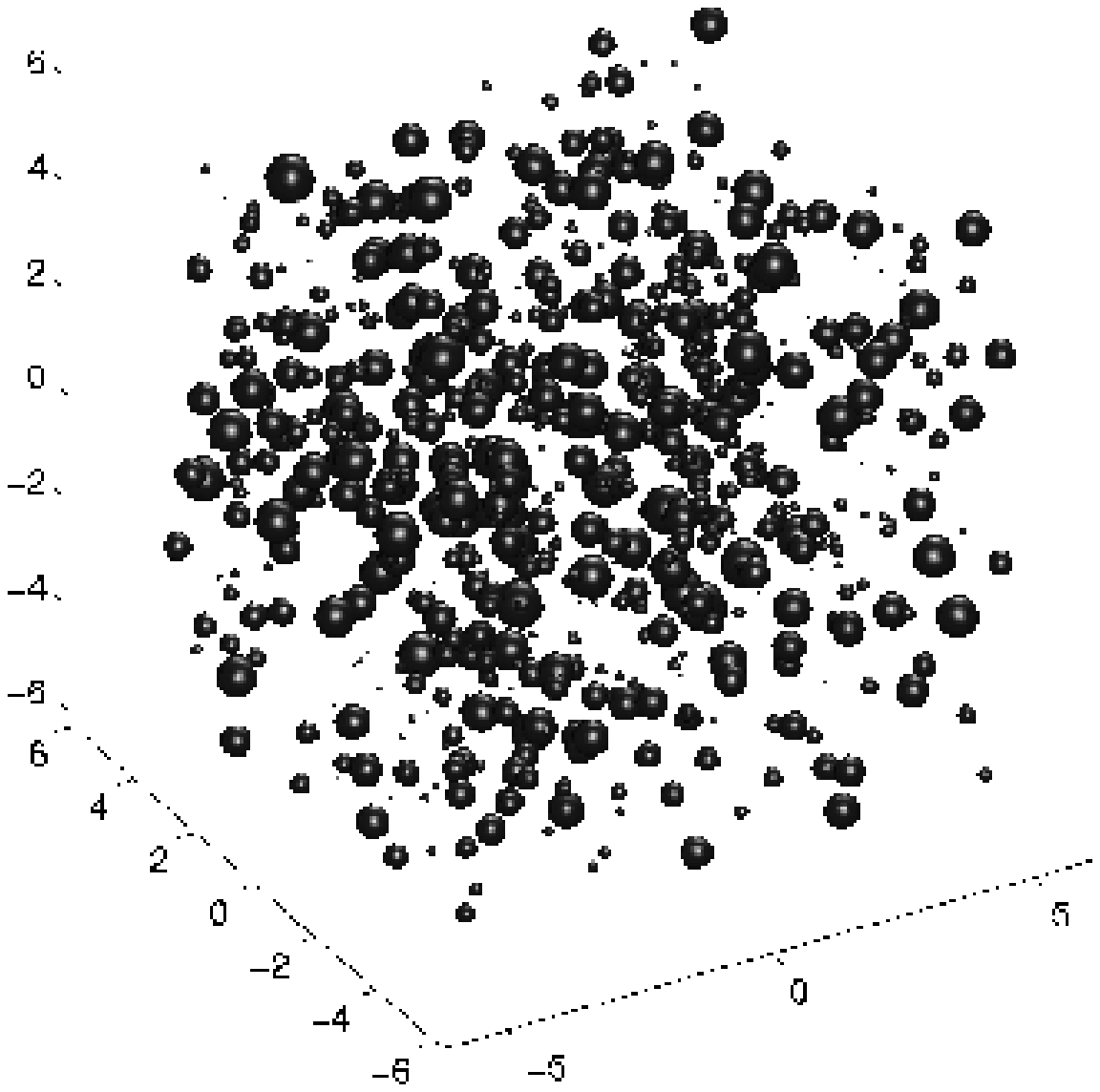,width=6.3cm}
\psfig{file=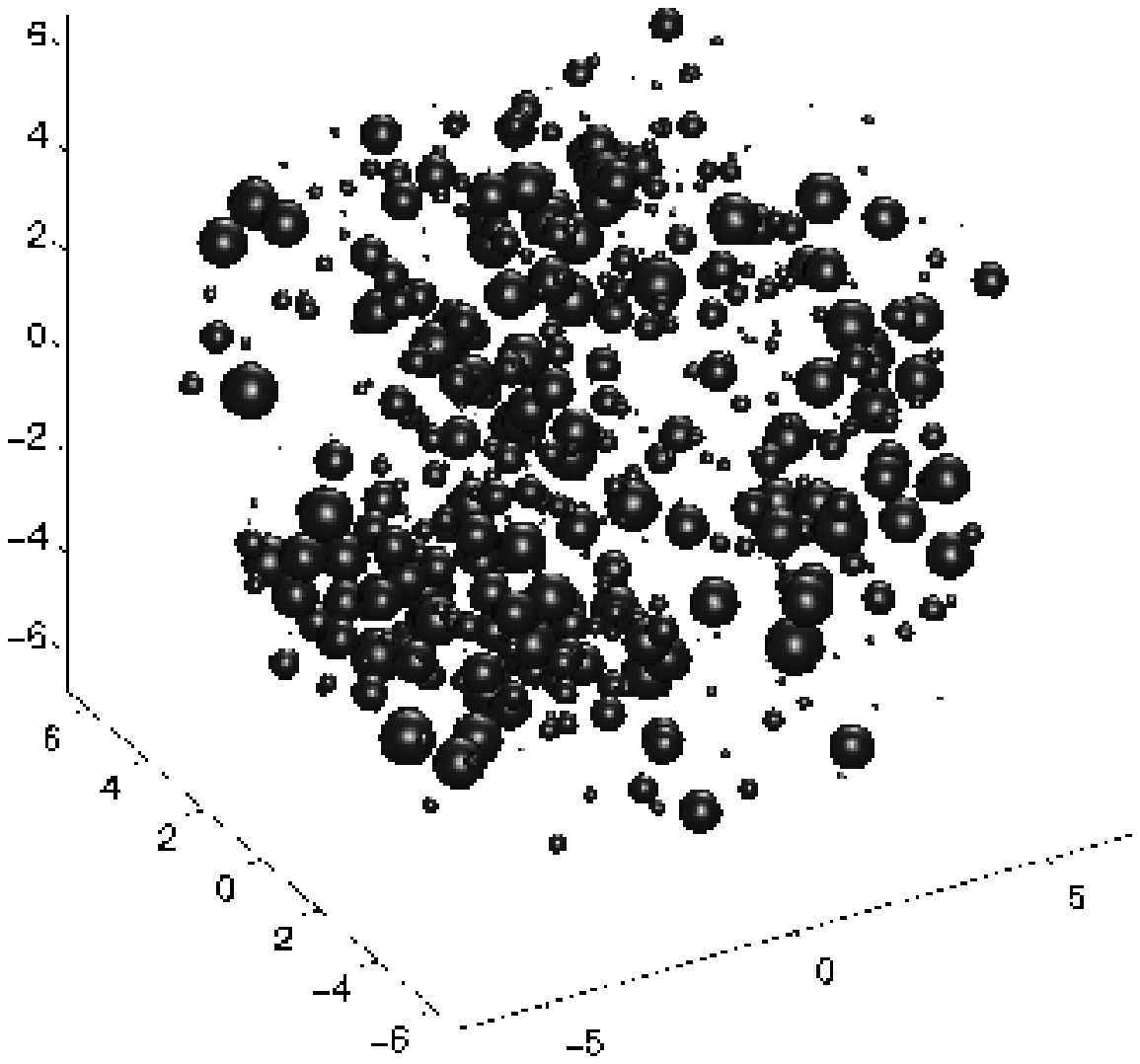,width=6.3cm}
\psfig{file=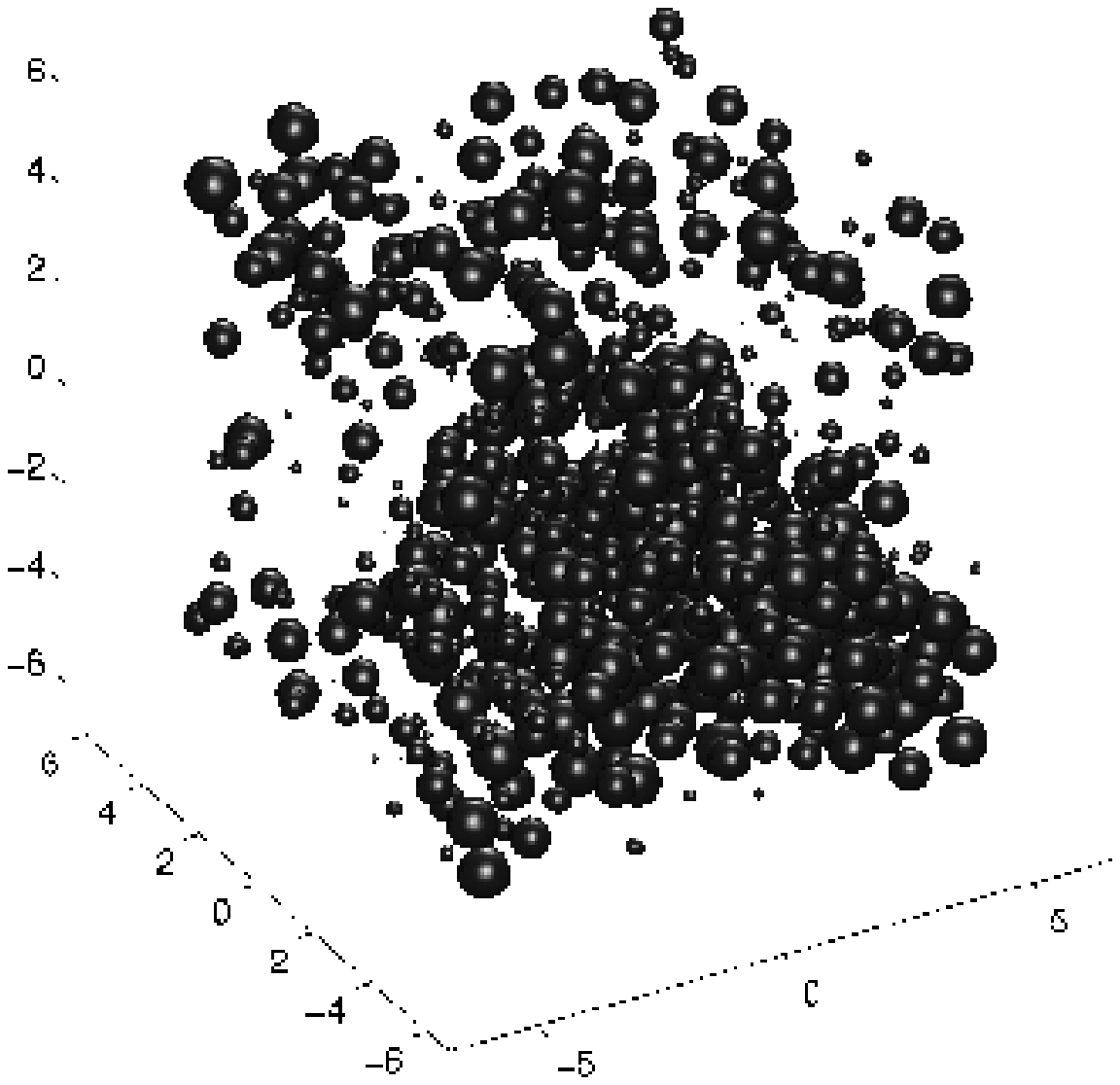,width=6.3cm}
\caption{\label{heterogeneity} 
Snapshots of the simulation box where points with
$\delta F = F_{\bf k}({\bf r},t) - F_s({\bf k},t) > 0$ 
are represented with radii proportional to $\delta F$ at temperatures
$T=2.0$, 0.6 and 0.45 (top to bottom), with $k=5.41$, $t = \tau(k=5.41,T)$.
Increasing spatial correlations of the particles' individual
dynamics when $T$ is lowered is evident.} 
\end{center}
\end{figure}

First, we measure the temperature dependent 
coherence length, $\ell(T)$,
associated with the ordering of the liquid's dynamics by measuring 
spatial correlations between individual particle 
relaxations~\cite{glo}. 
We use 
\be
F_{\bf k}({\bf r},t) = \sum_{j=1}^N \delta({\bf r}_j(0) -{\bf r}) 
\, \cos \big( {i {\bf k}
\cdot \left[ {\bf r}_j(t)-{\bf r}_j(0) \right] \big)}
\label{fkr}
\ee
as a natural local indicator of the dynamics, since 
$F_s({\bf k},t) = \langle F_{\bf k}({\bf r},t) \rangle$ 
is the real part of the standard self-intermediate scattering function;
$\langle \cdots \rangle$ stands 
for an ensemble average at temperature $T$, while
${\bf r}_j(t)$ is the position of particle $j$ at time $t$.
In our definition, dynamic heterogeneity implies that 
$F_{\bf k}({\bf r},t)$ becomes long-ranged 
correlated as $T$ is lowered~\cite{fss}, as can clearly be 
observed in the snapshots of Fig.~\ref{heterogeneity}. 

\begin{figure}
\begin{center}
\psfig{file=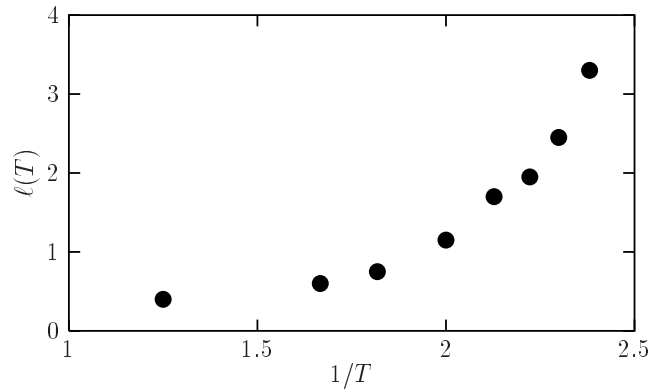,width=8.5cm}
\caption{\label{leng} 
Dependence of the coherence 
length $\ell(T)$ measuring the mean size of the clusters of 
Fig.~\ref{heterogeneity} on the inverse temperature from the
spatial fluctuations of $F_{\bf k}({\bf r},t)$, for $k=5.41$
and $t=\tau$, see Eq.~(\ref{cr}).
These data are discussed in detail in Ref.~\cite{steve}.}
\end{center}
\end{figure}

The measurement of the mean size of the dynamic clusters 
shown in Fig.~\ref{heterogeneity} involves the 
study of a two-point, two-time correlation 
function as already discussed in several
papers~\cite{juanpe,glo,steve,fss,reviewglo}. 
We extract $\ell(T)$ 
from the wavevector dependence of the Fourier transform of the 
correlator $C_{\bf k}({\bf r})$:
\be
C_{\bf k}
({\bf r}) = \frac{\langle F_{\bf k}({\bf 0},\tau) F_{\bf k}({\bf r},\tau) 
\rangle - 
\langle F_{\bf k}({\bf r},\tau) \rangle^2}
{\langle F_{\bf k}({\bf r},\tau)^2 \rangle - \langle F_{\bf k}({\bf r},\tau) 
\rangle^2},
\label{cr}
\ee
where $\tau=\tau(k = |{\bf k}|,T)$ is the relaxation time defined
in a standard way from the time decay 
of $F_s({\bf k},t)$~\cite{walter}.
The shape of $C_{\bf k}({\bf r})$, the 
temperature dependence of $\ell(T)$, and their 
theoretical interpretation in the context 
of a renormalization group analysis are the main 
object of Ref.~\cite{steve}, so that we only report the 
data for $\ell(T)$ in Fig.~\ref{leng} without discussing them further
in the present report.
Note that this first step is also the most demanding in terms of 
numerical resources.

\begin{figure}
\begin{center}
\psfig{file=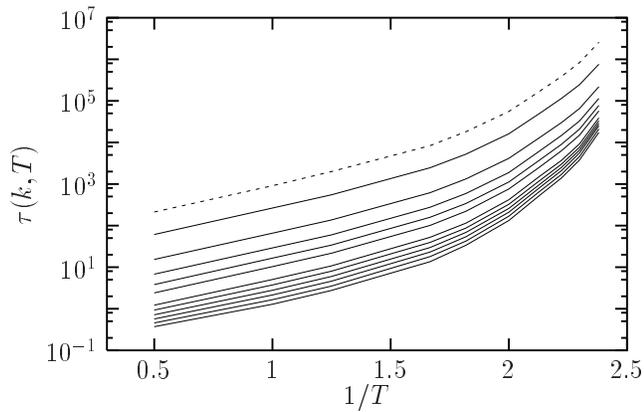,width=8.5cm}
\caption{\label{decoupling} Temperature dependence of 
$\tau(k,T)$ 
for various wavevectors, 
$k=7.21$, 6.61, 6.00, 5.41, 4.81, 4.21, 3.00, 2.40, 
1.80, 1.20, and 0.60 (bottom to top). The dashed line
is $10/D$, indicative of the $\tau(k \to 0,T)$ limit. Decoupling 
is observed as the temperature dependence is stronger 
at larger wavevectors.}
\end{center}
\end{figure}

Secondly, to probe decoupling, we 
measure with a great precision 
the temperature and wavevector 
dependences of the relaxation time $\tau(k,T)$
defined above. 
Our results are presented in Fig.~\ref{decoupling}.
We find that
the temperature dependence of $\tau(k \to 0,T)$ is 
the same as the inverse diffusion constant $D^{-1}$, as expected
in a diffusive regime, 
while for wavevectors close to the first peak 
of the static structure factor, $k_0 \approx 7.2$, the temperature
dependence is stronger, and follows that of the viscosity, 
establishing decoupling in our model system.
This finding has already been reported in several numerical 
works~\cite{jl,walter,glo,onuki}, and is not unexpected
in a system characterized by broad distributions of time scales.

We can now  establish our main result which is
the link between decoupling and dynamic heterogeneity. It  
stems from
intermediate wavevectors, $0 < k < k_0$, for which 
a crossover is observed in the temperature evolution
of $\tau(k,T)$, uniquely governed by the value 
of the scaling variable $k\ell(T)$. We find that
$\tau(k,T)$ follows $D^{-1}$ when $k \ell < 1$, 
or $\eta$ for $k \ell > 1$. 
It is therefore useful to define the following ratio, 
\be
X(k,T) = \frac{ \tau(k,T) D(T)}{\tau(k,T_o) D(T_o)}, 
\label{def}
\ee
since $X(k,T)$ is 
wavevector independent through the denominator, and is
temperature independent if $\tau(k,T) \propto D^{-1}(T)$, so that
by definition $X(k,T) = 1$ if no decoupling occurs.

\begin{figure}
\begin{center}
\psfig{file=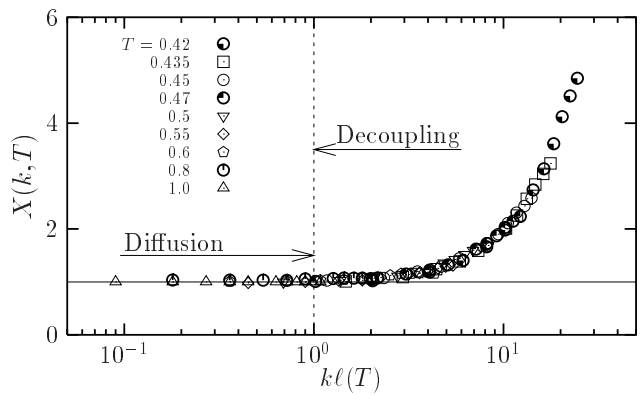,width=8.5cm}
\caption{\label{scaling}
The quantity $X(k,T)$, Eq.~(\ref{def}), 
as a function of the scaling variable 
$k \ell(T)$ for various $T$. The horizontal full line is the diffusive
prediction, $\tau D k^2 \sim const$. Departure from diffusion
arises at large $k\ell$, directly demonstrating
that the decoupling of different transport properties results
from an increasingly spatially correlated dynamics.}
\end{center}
\end{figure}

Instead we find that
time scales $\tau(k,T)$ spanning 7 orders of magnitude
can be collapsed on a unique, nontrivial curve, as shown in
Fig.~\ref{scaling}, so that
\be
X(k,T) \simeq {\cal X} [ k \ell(T) ].
\label{scal}
\ee
We find that ${\cal X}(x) = 1+x^\beta$ with 
$\beta \approx 1.6$ represents the numerical data quite well, 
so that the diffusive regime is limited to small 
values of the scaling variable $k\ell$.
Note that this scaling extends from the onset temperature $T_o$ to well below 
the mode-coupling temperature $T_c$.
From the scaling behaviour (\ref{scal}), 
our main conclusion is therefore that 
dynamic heterogeneity emerging at $T_o$ 
and increasing when $T$ is lowered
is directly responsible for decoupling.

Decoupling phenomena have been experimentally characterized close 
to the calorimetric glass transition where they are more pronounced, 
as can be understood from Fig.~\ref{scaling}. We believe, however, 
that the behaviour is qualitatively similar to our numerical
investigations because even 
an increase of several decades in time scales corresponds to a very modest
change of the coherence length $\ell(T)$, 
which is the key quantity of the problem. 
Experiments report moreover that $D \eta \sim \eta^{\alpha}$ 
with $1 > \alpha > 0$~\cite{zphys,exp,hetero}.
We find similarly that, at fixed $k$,
$\tau D \sim \ell^\beta$. 
The dynamic scaling 
discussed in Ref.~\cite{steve}, $\ell \sim \tau^{1/z}$, yields indeed
the observed power law,
\begin{equation}
D \eta \sim \eta^{\beta/z},
\label{expe}
\end{equation} 
with an exponent $\beta/z \approx 0.35$. 
Various values for this exponent have been reported
from experiments~\cite{hetero,andreozzi} and
a precise characterization of all these exponents for various
liquids on a wide range of length scales 
would be most useful.
Note that the use of the crossover scaling function
${\cal X}(x)$ in Eq.~(\ref{expe}) would automatically 
yield a smaller effective value of the exponent $\beta/z$ on a restricted time
window.

Finally, we discuss the theoretical interpretations and consequences
of Fig.~\ref{scaling}.
These results constitute the first direct demonstration
that, much as in ordinary critical phenomena, large spatial correlations 
not only accompany but also influence the properties of 
supercooled liquids, and therefore the glass formation itself.
Thus, they are a striking confirmation that spatial 
approaches are necessary to understand the formation of glasses, and 
several such quantitative approaches can be found in the 
literature~\cite{steve,0T,juanpe2,tarjus,wolynes}.

When a growing length scale for dynamic heteroheneity is 
included, our results are very naturally explained, just as  
the decoupling phenomenon is~\cite{theo}. Tuning
the wavector in (\ref{fkr}) amounts to probing the dynamics 
on different length scales, smoothly interpolating between 
$D^{-1}$ ($k \to 0$) and $\eta$ ($k\approx k_0$), as is 
observed in Fig.~\ref{scaling} which clearly demonstrates
that the crossover, $k\ell(T) \sim 1$, is ruled by the increasing correlation 
length of dynamic heterogeneity.

Our findings also confirm that dynamic heterogeneity is 
a central aspect of the dynamics of supercooled liquids in that
time and length scales are intimately connected~\cite{juanpe,0T}. 
This is at odds with the opposite belief that slow dynamics
emerges because of the local blocking of the particles without
any relevant length scale beyond the interparticle distance, 
the famous ``cage effect''~\cite{MCT}.
Our results make it clear that the alpha-relaxation is 
instead a cooperative
phenomenon where single particle dynamics are coherent on the length 
scale $\ell(T)$ much larger than $k_0^{-1}$, which is in turn directly 
responsible for the temperature behaviour of the time scales.
Moreover, $\ell(T)$ starts to grow significantly
and connects to time scales even in the regime $T_c < T < T_o$
where a mode-coupling analysis supposedly applies~\cite{walter,MCT}.
This confirms that heterogeneous dynamics, decoupling and 
activated dynamics set in at the onset temperature $T_o$ which 
is therefore the key temperature scale of the problem, as opposed to 
$T_c$ where no significant change of mechanism takes 
place~\cite{0T,reich}. We note that
the absence of non-trivial spatial correlations~\cite{franz}, the incorrect
identification of $T_c$ as a key temperature scale, and the absence of 
decoupling in the present formulation of the mode-coupling theory 
represent major failures of this approach.

Our results constitute therefore 
a sharp new test to discriminate between the 
many theoretical approaches to the glass transition
problem~\cite{review}.
Indeed, any theory in which
time scales do not directly follow from the existence
of spatial correlations growing when $T$ is decreased below the
onset temperature $T_o$ is seriously challenged by the present work. 
Quite importantly, this includes a number 
of approaches which have been nonetheless much applied 
to describe experimental results~\cite{free,MCT,landscape}.

Our conclusions are drawn from a scaling relation discovered for a specific, 
yet paradigmatic, model system. Although we believe they are 
generic to supercooled liquids, the absence 
of any data on different systems emphasizes both the novelty of this work and 
the need for further detailed investigations
of the slow dynamics of various supercooled liquids. 
It is clear, for instance,
that the precise experimental characterization of $\ell(T)$ should
become a central goal for future investigations, and 
the present work therefore suggests a new way to 
access physically relevant length scales. 

I thank S. Whitelam and J.P. Garrahan for a
fruitful collaboration~\cite{steve} and discussions,
J.-L. Barrat,  J.-P. Bouchaud and G. Tarjus for useful correspondence. 
This work is supported by CNRS (France), E.U. (Marie Curie
Grant No.\ HPMF-CT-2002-01927), Worcester College Oxford (UK), and 
Oxford Supercomputing Centre at Oxford University (UK).

\end{document}